\documentclass[preprint2]{aastex}

\newcommand{\vsini}{\mbox{$v \sin I$}}
\newcommand{\feh}{\mbox{[Fe/H]}}
\newcommand{\teff}{\mbox{$T_{\rm eff}$}}
\def\secos{$\sqrt{e} \cos \omega$}
\def\sesin{$\sqrt{e} \sin \omega$}
\def\qps{$Q^{\prime}_{\rm *}$}
\def\qpp{$Q^{\prime}_{\rm P}$}
\def\etal{et\,al.}
\def\sqiglt{\hbox{\rlap{\lower.55ex \hbox {$\sim$}}
        \kern-.3em \raise.4ex \hbox{$<$}\,}}
\def\sqiggt{\hbox{\rlap{\lower.55ex \hbox {$\sim$}}
        \kern-.3em \raise.4ex \hbox{$>$}\,}}

\slugcomment{To appear in Astrophysical Journal Letters}

\shorttitle{The orbit of WASP-19b}
\shortauthors{Hellier et al.}

\begin{document}

\title{On the orbit of the short-period exoplanet WASP-19b}

\author{
Coel Hellier\altaffilmark{1}, 
D.R. Anderson\altaffilmark{1}, 
A. Collier Cameron\altaffilmark{2}, 
G.R.M. Miller\altaffilmark{2},
D. Queloz\altaffilmark{3}, 
B. Smalley\altaffilmark{1}, 
J. Southworth\altaffilmark{1} \&\ 
A.H.M.J. Triaud\altaffilmark{3}.
}

\altaffiltext{1}{Astrophysics Group, Keele University, Staffordshire, ST5 5BG, UK}
\altaffiltext{2}{SUPA, School of Physics and Astronomy, University of St.\ Andrews, North Haugh,  Fife, KY16 9SS, UK}
\altaffiltext{3}{Observatoire astronomique de l'Universit\'e de Gen\`eve
51 ch. des Maillettes, 1290 Sauverny, Switzerland}

\begin{abstract}
WASP-19b has the shortest orbital period of any known exoplanet, 
orbiting at only 1.2 times the Roche tidal radius.  By observing
the Rossiter--McLaughlin effect we show that WASP-19b's orbit
is aligned, with $\lambda =  4.6 \pm 5.2^{\circ}$.  Using, in addition,  
a spectroscopic \vsini\ and the observed rotation period
we conclude that the obliquity, $\psi$, is less than 
$20^{\circ}$. Further, the eccentricity of the orbit is less 
than 0.02.  We argue that hot Jupiters with orbital periods
as short as that of WASP-19b are two orders of magnitude less
common than hot Jupiters at the 3--4-d `pileup'. We 
discuss the evolution of WASP-19b's orbit and 
argue that most likely it was first moved to near twice the Roche 
limit by third-body interactions, and has since spiralled 
inwards to its current location under tidal decay.  This is
compatible with a stellar
tidal-dissipation quality factor, \qps, of order 10$^{7}$.  
\end{abstract}


\section{Introduction}
Of the known transiting exoplanets WASP-19b has the shortest
orbital period, at only 0.79 d, and thus is important for
constraining models of the orbital evolution of hot-Jupiter
planets.  This area of theory is currently undergoing rapid
development owing to the finding of planets
with apparently short dynamical lifetimes owing to  
strong tidal interactions with their host stars (e.g.\ Rasio \etal\
1996; Levrard, Winisdoerffer \&\ Chabrier 2009; Matsumura, Peale \&\
Rasio 2010; Penev \&\ Sasselov 2011), of which 
WASP-18b (Hellier \etal\ 2009) and WASP-19b (Hebb \etal\ 2010) 
are among the most extreme. 

Further, many of the orbits of hot Jupiters are being found to be 
misaligned with the stellar spin axis, with some being 
retrograde (e.g.\ Winn \etal\ 2009; Narita \etal\ 2009; Triaud \etal\ 2010),
which cannot be accounted for solely by the `migration' 
scenario (e.g.\ Lin \etal\ 1996) in which hot Jupiters form 
further out and migrate through a protoplanetary disc to 
short-period orbits.

Thus a picture is emerging in which hot-Jupiter orbits result from a
mixture of processes including: disc migration; planet--planet
scattering and the Kozai mechanism, by which planets can be driven
into eccentric or misaligned orbits through the influence of a third
perturbing body; and tidal dissipation, which circularises and
shortens orbits (e.g.\ Fabrycky \&\ Tremaine 2007; Nagasawa, Ida \&\
Bessho 2008; Matsumura \etal\ 2010; Naoz \etal\ 2010).

The relative importance of
these mechanisms for the ensemble of hot Jupiters 
can be investigated particularly
by using the transiting exoplanets (e.g.\ Triaud \etal\ 2010; 
Morton \&\ Johnson 2011), since the 
Rossiter--McLaughlin effect (a distortion of the star's
line profiles as the planet transits the stellar disc) 
tells us the angle between the planetary orbit and the 
sky-projection of the stellar spin axis.

Ford \&\ Rasio (2006) pointed out that the orbital period
distribution of hot Jupiters cuts off near 2 Roche radii, 
which could be explained  if planets are 
scattered into highly elliptical orbits from much further out 
and then circularise while conserving angular momentum 
(in contrast, disc migration alone would
predict a smooth distribution down to the Roche limit). 

WASP-19b is one of a small number of planets with substantially
smaller orbital radii, and is the most extreme at only 1.2 times 
the Roche limit. Guillochon, Ramirez-Ruiz \&\ Lin (2010) argue 
that it would be hard to scatter a planet 
into such an orbit, since it  would instead by destroyed 
or ejected, and argue that either disc migration
prior to scattering or tidal decay of the orbit after scattering
is also necessary. 

We report here an observation of the Rossiter--McLaughlin 
(R--M) effect for WASP-19b, which we analyse along with other data
to further constrain WASP-19b's orbit, and  discuss how
this planet fits into the above theoretical picture. 

\begin{table}
\caption{Radial velocity measurements of WASP-19. The first
three are from CORALIE and the remainder from HARPS.} 
\label{rv-data} 
\begin{tabular*}{0.5\textwidth}{@{\extracolsep{\fill}}cccc} 
\hline 
BJD--2\,400\,000 & RV & $\sigma$$_{\rm RV}$ & BS\\ 
 & (km s$^{-1}$) & (km s$^{-1}$) & (km s$^{-1}$)\\ 
\hline
54972.4982 & 20.8874 & 0.0176 & \rule{0mm}{5mm}0.0066\\ [-0.0mm]
54973.4648 & 20.6042 & 0.0182 & $\!\!$--0.0806\\ [-0.0mm]
54999.4857 & 20.5571 & 0.0341 & 0.0223\\ [-0.0mm]
55242.6884 & 20.7469 & 0.0036 & $\!\!$--0.0074\\ [-0.0mm]
55242.8457 & 21.0203 & 0.0050 & $\!\!$--0.0197\\ [-0.0mm]
55243.6590 & 21.0391 & 0.0042 & 0.0116\\ [-0.0mm]
55243.8242 & 20.9547 & 0.0043 & 0.0213\\ [-0.0mm]
55244.7221 & 20.7355 & 0.0031 & 0.0081\\ [-0.0mm]
55247.7151 & 21.0254 & 0.0088 & 0.0202\\ [-0.0mm]
55272.5445 & 20.5741 & 0.0044 & $\!\!$--0.0180\\ [-0.0mm]
55272.7776 & 20.9778 & 0.0052 & $\!\!$--0.0280\\ [-0.0mm]
55274.5328 & 21.0391 & 0.0073 & 0.0056\\ [-0.0mm]
55274.6147 & 20.9318 & 0.0124 & $\!\!$--0.0369\\ [-0.0mm]
55274.6224 & 20.9129 & 0.0112 & $\!\!$--0.0573\\ [-0.0mm]
55274.6299 & 20.8819 & 0.0094 & $\!\!$--0.0486\\ [-0.0mm]
55274.6375 & 20.8842 & 0.0088 & 0.0044\\ [-0.0mm]
55274.6451 & 20.8711 & 0.0084 & $\!\!$--0.0122\\ [-0.0mm]
55274.6525 & 20.8914 & 0.0086 & 0.0135\\ [-0.0mm]
55274.6602 & 20.8473 & 0.0094 & $\!\!$--0.0158\\ [-0.0mm]
55274.6679 & 20.8214 & 0.0100 & 0.0340\\ [-0.0mm]
55274.6754 & 20.7819 & 0.0108 & 0.0141\\ [-0.0mm]
55274.6830 & 20.7568 & 0.0118 & 0.0626\\ [-0.0mm]
55274.6907 & 20.7315 & 0.0108 & $\!\!$--0.0030\\ [-0.0mm]
55274.6983 & 20.7138 & 0.0103 & 0.0136\\ [-0.0mm]
55274.7058 & 20.7262 & 0.0106 & 0.0147\\ [-0.0mm]
55274.7134 & 20.7093 & 0.0104 & $\!\!$--0.0431\\ [-0.0mm]
55274.7210 & 20.7151 & 0.0118 & $\!\!$--0.0090\\ [-0.0mm]
55274.7288 & 20.7056 & 0.0120 & $\!\!$--0.0379\\ [-0.0mm]
55274.7363 & 20.6582 & 0.0116 & 0.0071\\ [-0.0mm]
55274.7438 & 20.6687 & 0.0104 & $\!\!$--0.0343\\ [-0.0mm]
55274.7514 & 20.6387 & 0.0119 & 0.0095\\ [-0.0mm]
55274.7591 & 20.6452 & 0.0113 & $\!\!$--0.0058\\ [-0.0mm]
55274.7667 & 20.6136 & 0.0105 & $\!\!$--0.0057\\ [-0.0mm]
55274.7740 & 20.6079 & 0.0106 & $\!\!$--0.0005\\ [-0.0mm]
55274.7818 & 20.5771 & 0.0123 & 0.0043\\ [-0.0mm]
55274.8216 & 20.5707 & 0.0069 & $\!\!$--0.0187\\ [-0.0mm]
55275.5221 & 20.6841 & 0.0052 & $\!\!$--0.0226\\ [-0.0mm]
55275.7895 & 20.6841 & 0.0077 & $\!\!$--0.0016\\ [-0.0mm]
55276.5146 & 20.5850 & 0.0062 & 0.0018\\ [-0.0mm]
\hline  
\multicolumn{4}{l}{Bisector errors are twice RV errors} 
\end{tabular*} 
\end{table}

\section{Observations}
We obtained 25 spectra of WASP-19 through a transit
on 2010 March 19 using the HARPS spectrograph on the ESO
3.6-m at La Silla.   The data were reduced using the
Data Reduction Software by removing the blaze function
and cross-correlating with a G2 mask 
(see Triaud \etal\ 2010 for further details).  In addition
2 HARPS spectra were taken two days earlier and 3 more
over the next two days to better constrain
the in-transit data.  We also report 6 more 
HARPS spectra and 3 spectra from the CORALIE spectrograph
on the Swiss Euler 1.2-m telescope obtained to tie down the 
eccentricity of the orbit (the 39 new radial velocities
are listed in Table~1).  In the following analysis we
also include the 34 CORALIE radial velocities previously
listed by Hebb \etal\ (2010). 

To constrain the transit time, duration and ingress and egress
lengths, thus improving the R--M analysis, we included a 
photometric Gunn $r$-band transit obtained using the 
ESO NTT/EFOSC on the night of 2010 February 28, 19 days from 
the HARPS transit. The telescope was defocused so that each PSF 
covered approximately 8500 pixels. The data were reduced 
using aperture photometry relative to an optimal combination 
of six comparison stars following the methods of 
Southworth \etal\ (2009).  We also included 17\,162 WASP-South 
photometric data points from 2006--2008 and the FTS $z$-band 
lightcurve reported by Hebb \etal\ (2010). 

The time of secondary occultation (of the planet by the star)
helps to tie down the eccentricity, and thus we included
in our analysis the $H$-band occultation photometry reported
by Anderson \etal\ (2010), detrending it with a quadratic
function of time and sky-background, as in that paper, and 
also the $K$-band data of Gibson \etal\ (2010), with a linear 
detrending in time.

\begin{figure}
\hspace*{-5mm}\includegraphics[width=9cm]{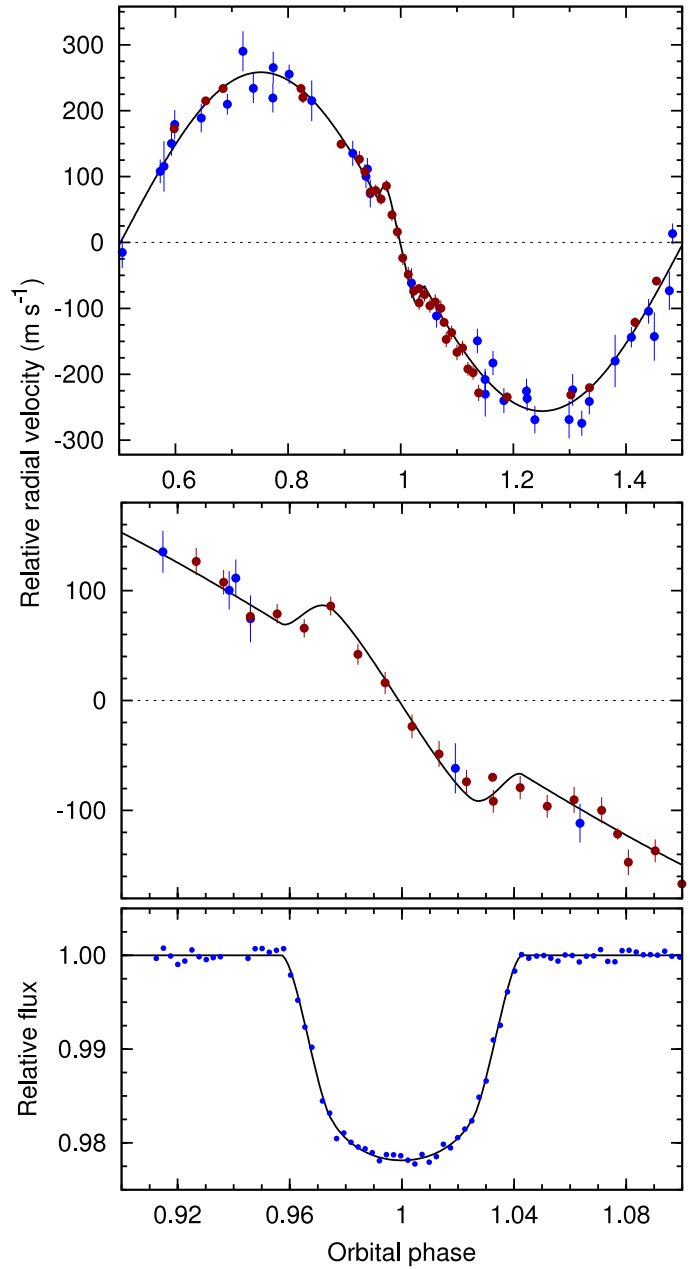}\\ [-2mm]
\caption{(Top) The HARPS (brown) and CORALIE (blue) radial 
velocities of WASP-19
together with the fitted model. (Second panel) The transit region shown
expanded. (Lower panel) The NTT transit lightcurve and fitted model.}
\end{figure}

\section{Analysis}
To model WASP-19b's orbit we used a Markov Chain Monte Carlo (MCMC)
approach similar to that used by Triaud \etal\ (2010) and based on an
R--M model code described more fully in Triaud \etal\ (2009).  The datasets
listed in the previous section were fitted simultaneously with a model
based on the parameters $T_{\rm c}$, $P$, $\Delta F$, $T_{14}$, $b$,
$K_{\rm 1}$, \teff, \feh, \secos, \sesin, $\sqrt{\vsini} \cos
\lambda$, $\sqrt{\vsini} \sin \lambda$,and, for the occultation data,
$\Delta F_{1.6\,\micron}$ and $\Delta F_{2.09\,\micron}$.  Here,
$T_{\rm c}$ is the epoch of mid-transit, $P$ is the orbital period,
$\Delta F$ is the fractional flux-deficit that would be observed
during transit in the absence of limb-darkening, $T_{14}$ is the total
transit duration (from first to fourth contact), $b$ is the impact
parameter of the planet's path across the stellar disc, $K_{\rm 1}$ is
the stellar reflex velocity semi-amplitude, \teff\ is the stellar
effective temperature, \feh\ is the stellar metallicity, $e$ is the
orbital eccentricity, $\omega$ is the argument of periastron, and
$\Delta F_{1.6\,\micron}$ and $\Delta F_{2.09\,\micron}$ are the
planet/star flux ratios at 1.6~\micron\ and 2.09~\micron\
respectively.  The resulting system parameters are listed in Table~2
and the model fits are illustrated in Figure~1.

As inputs to the MCMC we used Gaussian ``priors'' of \teff\ =
5500\,$\pm$\,100 K and [Fe/H] = 0.02\,$\pm$\,0.09 (from Hebb \etal\
2010).  We also applied a prior on \vsini\ of 5.0\,$\pm$\,0.3
km\,s$^{-1}$ (where we use \vsini\ for the projected stellar rotation
to distinguish from $i$ as the inclination of the planet's orbit).
This was estimated from the HARPS spectra by fitting the profiles of
several unblended Fe\,{\sc i} lines. We measured an
instrumental broadening of 0.06 \AA\ from telluric lines and we
assumed a value for macroturbulence of 1.7\,$\pm$\,0.3 km\,s$^{-1}$
(Bruntt \etal\ 2010).

As discussed by Triaud \etal\ (2011) the choice of
\vsini\ prior is critical in modelling the R--M effect when the
planet has a low impact parameter.  This is because an orbit
at right-angles to the spin axis and with zero impact parameter
is insensitive to \vsini.  However, for WASP-19b the impact
parameter ($b$ = 0.66\,$\pm$\,0.02) is high enough to remove
the degeneracy, and omitting the prior makes little difference
(the R--M angle, $\lambda$, is 4.6\,$\pm$\,5.2$^{\circ}$ with
the \vsini\ prior and 6.2\,$\pm$\,6.8$^{\circ}$ without).   

The stellar \teff, along with the stellar density and metallicity,
was propagated to the stellar mass using
the calibration of Enoch \etal\ (2010). \teff\ was also used to 
select limb-darkening coefficients from the four-parameter
models listed by Claret (2000; 2004). We also added jitter
of  14.1 m\,s$^{-1}$ to the CORALIE radial velocities
and  6.9 m\,s$^{-1}$ to the HARPS radial velocities in order 
to obtain a spectroscopic 
$\chi^{2}_{\nu}$ of 1 during the fit and thus balance the 
different datasets in the MCMC.  

The resulting fit (Fig.~1) has a $\chi^{2}$ of 176 ($\nu$ = 68)
for the radial-velocity data (calculated without added jitter), 
which likely indicates that stellar activity is affecting the
radial-velocities, and that there are non-uniformities on
the face of the star during the transit. Note that WASP-19
has shown star spots during transits (Southworth, unpublished
data), that the star shows a rotational photometric modulation 
at 10.5 d (Hebb \etal\ 2010), and also that the most discrepant 
HARPS point during transit (at phase 1.032) was taken 31 days 
prior to the main HARPS transit.  Further, the size of 
the residuals is in line with that expected from activity 
in spotted G stars (Saar \&\ Donahue 1997). The fit to the 
NTT transit lightcurve has a $\chi^{2}$ of 69 ($\nu$ = 70).

\begin{table}[t]
\caption{System parameters for WASP-19.\protect\rule[-1.5mm]{0mm}{2mm}}  
\begin{tabular}{lc}
Parameter (Unit) & Value \rule{0mm}{5mm} \\ [0.5mm] 
\hline
$P$ (d) & 0.7888400 $\pm$ 0.0000003\\
$T_{\rm c}$ (HJD) & 2455168.96801 $\pm$ 0.00009 \\
$T_{\rm 14}$ (d) & 0.0655 $\pm$ 0.0003 \\
$T_{\rm 12}=T_{\rm 34}$ (d) & 0.0135 $\pm$ 0.0005 \\
$\Delta F=R_{\rm P}^{2}$/R$_{*}^{2}$ & 0.0206 $\pm$ 0.0002\\
$b$ & 0.657 $\pm$ 0.015\\ 
$i$ ($^\circ$)  & 79.4 $\pm$ 0.4\\
$K_{\rm 1}$ (km s$^{-1}$) & 0.257 $\pm$ 0.003\\
$\gamma$ (km s$^{-1}$) & 20.7873 $\pm$ 0.0002\\
$e\cos\omega$ & 0.0024 $\pm$ 0.0020\\ 
$e\sin\omega$ & 0.000 $\pm$ 0.005\\ 
$e$ & 0.0046$^{+ 0.0044}_{- 0.0028}$\\
  &  $<$\,0.02 (3$\sigma$)\\ 
$\omega$ ($^\circ$) & 3 $\pm$ 70\\ 
$\phi_{\rm mid-occult}$ & 0.5015 $\pm$ 0.0012\\
$v \sin I$ (km s$^{-1}$) & 4.63 $\pm$ 0.26\\
$\lambda$ ($^\circ$) & 4.6 $\pm$ 5.2\\
$\psi$ ($^\circ$) & $<$\,20 (2$\sigma$)\\
$M_{\rm *}$ (M$_{\rm \odot}$) & 0.97 $\pm$ 0.02\\
$R_{\rm *}$ (R$_{\rm \odot}$) & 0.99 $\pm$ 0.02\\
$\log g_{*}$ (cgs) & 4.432 $\pm$ 0.013\\
$\rho_{\rm *}$ ($\rho_{\rm \odot}$)  & 0.993$^{+ 0.047}_{- 0.042}$\\
$M_{\rm P}$ (M$_{\rm Jup}$) & 1.168 $\pm$ 0.023\\
$R_{\rm P}$ (R$_{\rm Jup}$) & 1.386 $\pm$ 0.032\\
$\log g_{\rm P}$ (cgs) & 3.143 $\pm$ 0.018\\
$\rho_{\rm P}$ ($\rho_{\rm J}$) & 0.438 $\pm$ 0.028\\
$a$ (AU)  & 0.01655 $\pm$ 0.00013\\
$T_{\rm P, A=0}$ (K) & 2050 $\pm$ 40\\
\hline
\end{tabular} 
\end{table} 

\section{Results and Discussion} 
We find that WASP-19b is in an aligned orbit, with $\lambda = 4.6 \pm\
5.2^{\circ}$, where $\lambda$ is the angle between the planet's orbit
and the sky-projected stellar rotation axis.  Further we find that
the eccentricity is compatible with zero, with a 1-$\sigma$ upper 
limit of 0.009 and a 3-$\sigma$ upper limit of 0.02.    

Hebb \etal\ (2010) reported a
rotational modulation in the WASP data at a period of 10.5\,$\pm$\,0.2
d. With the stellar radius from Table~2 this would translate to an
equatorial velocity of 4.77\,$\pm$\,0.13 km\,s$^{-1}$. This estimate of
$v$ is compatible both with our spectroscopic \vsini\ of
5.0\,$\pm$\,0.3  km\,s$^{-1}$  and with the \vsini\ fitted to the R--M of
4.6\,$\pm$\,0.3  km\,s$^{-1}$. That match implies that the stellar spin axis is
nearly perpendicular to the line of sight, which suggests that not only is
 $\lambda$ low but that the obliquity itself, $\psi$, is also low. 
Since the impact factor ($b$ = 0.66) is relatively
high, the R--M value will sample higher latitudes and thus might 
be reduced by differential rotation. We therefore take the 
spectroscopic \vsini\ as
a better indicator of the projected equatorial velocity.  That results
in a 2$\sigma$ limit of $I > 65^{\circ}$, which translates 
(using information from Table 2 and equation 9 from Fabrycky 
\&\ Winn 2009) to a 2$\sigma$ upper limit of $\psi < 20^{\circ}$.

The known hot Jupiter exoplanets show a `pileup' around periods of 3--4 d
(e.g.\ Szab\'o \&\ Kiss 2011).  The smaller number at longer periods
($\sim$\,10 d) is to some extent a selection effect, resulting
from the limited time sampling of ground-based transit surveys. 
The candidate list for Jupiter-sized objects from the better-sampled
{\sl Kepler\/} data  shows proportionately more at
longer periods, but still shows a peak at 3--4 d and a decline beyond
that (Borucki \etal\ 2011). The 
reduced number at periods below 2 d is definitely 
real, being seen in both ground- and space-based transit surveys and in
radial-velocity surveys. For example the frequency of Jupiter-sized 
{\sl Kepler\/} candidates drops by more than an order of magnitude 
below semi-major axis $\sim$ 0.03 AU (period $\sim$ 2 d) 
(Borucki \etal\ 2011), and note that this is a lower limit to the
fall-off of planets, since as the planets become rarer the transit-mimics 
would become a larger fraction of the candidates.  

Similarly, the WASP-South survey, which has
found the 3 known Jupiter-sized planets with periods $<$\,1 d 
(WASP-18b, Hellier \etal\ 2009; WASP-19b, Hebb \etal\ 2010; WASP-43b,
Hellier \etal\ 2011),  shows a much reduced  `hit rate' for planets 
below 2 d (Hellier \etal\ 2010), even though it will be
most sensitive to such planets, since they produce the most transits 
and also because the transit durations are shorter compared to the 
data lengths.  Thus the transiting Jupiters with periods below 
1 d are rare, and found only because the WASP survey covers more than 
an order of magnitude more stars than {\sl Kepler}.  

Note, further,
that this means that the actual frequency of such planets
(not just transiting ones) will be lower still, since the range
of inclinations that produces a transit rises steeply for decreasing
semi-major axis for such close-in planets. For example WASP-19b would transit
for inclinations $>$\,74$^{\circ}$, but, if it were in the pileup near
3--4 d, the limit would be $\sim$\,85$^{\circ}$, a factor 3 reduction in 
probability.  Thus a tentative estimate is that planets such as 
WASP-19b are two orders of magnitude less common than hot-Jupiters at 
3--4 d. This means that either Jupiter-sized planets cannot
easily get into such orbits, or that something, such as tidal
orbital decay, is then rapidly destroying them.

Ford \&\ Rasio (2006) noted that the lower edge of the hot-Jupiter 
pileup was near twice the Roche limit, whereas disc 
migration would be
expected to produce a smooth distribution down to the Roche
limit. They argued that this would arise naturally if many
hot Jupiters arrived from much further out by being scattered 
into highly eccentric orbits that were then circularised
while conserving angular momentum.  
The finding of many misaligned and retrograde planet orbits
among those in the pileup (Triaud \etal\ 2010; Winn \etal\ 2010; 
Narita \etal\ 2010) also argues that these systems (or a large
fraction of them) arise not from simple migration, but from
third-body process such as planet--planet scattering or the 
Kozai mechanism (e.g.\ Fabrycky \& Tremaine 2007; Nagasawa \etal\ 2008; 
Triaud \etal\ 2010). 

However, there are 5 hot Jupiters further in than 2
Roche radii (see Matsumura \etal\ 2010), of which WASP-19b is 
the most extreme.  Using the
parameters of Table~2, WASP-19b, with an orbital period of 0.79 d, has
a semi-major axis of only 1.21 times the Roche tidal radius ($a_{\rm
R} \approx\ 2.16 R_{\rm P} (M_{\rm *}/M_{\rm P})^{1/3})$.
Guillochon \etal\ (2010) argue that such orbits are unlikely
to result from the scattering of planets from outside the ice line, 
followed by circularisation,  since the planets would instead by 
destroyed or ejected. Thus they argue that these planets must have
migrated inwards prior to scattering, or have spiralled inwards after 
scattering as a result of tidal decay.  

Given that the tidal decay timescale for WASP-19b
in its current orbit is likely to be significantly shorter than the 
system age, we suggest, following Matsumura \etal\ (2010) and 
Guillochon \etal\ (2010), that the most likely scenario for
WASP-19b is: formation beyond the ice line; transfer to an
orbit near 2\,$a_{\rm R}$, at the short-period edge of the 
hot-Jupiter `pileup', by scattering or by the Kozai
mechanism, followed by circularisation at 2\,$a_{\rm R}$; 
and then tidal decay of the orbit to the current 1.2\,$a_{\rm R}$. 

The timescale for tidal decay of the orbit is set
by the tidal dissipation in the star, denoted by the 
stellar quality factor \qps, while the eccentricity
damping is also affected by the planetary dissipation,
\qpp, and so could proceed faster, allowing  
circularisation at 2\,$a_{\rm R}$ before significant
decay of the orbit (Matsumura \etal\ 2010). 
To investigate whether this scenario is consistent with the
observed parameters of WASP-19b we need a value
of \qps\ small enough such that WASP-19b's orbit will decay
from $\sim$\,2\,$a_{\rm R}$ to 1.2\,$a_{\rm R}$ within the lifetime
of WASP-19, estimated at \sqiggt 1 Gyr by Hebb \etal\ (2010),
yet large enough to result in a long-enough lifetime of WASP-19b
to give a reasonable probability of now observing it.  

Using eqn 5 of Levrard \etal\ (2009)\footnote{See Matsumura
\etal\ (2010) for corrections to some of the equations in this paper.}
we find that to
have decayed inwards from 2\,$a_{\rm R}$ within 1 Gyr would
require a \qps\ no higher than $10^{7}$, while a 10-Gyr
age would allow up to $10^{8}$ (though note that we also
need time for the third-body interactions leading to the 
starting point of $\sim$\,2$a_{\rm R}$).  Using the same equation
the remaining lifetime would be 40 Myr for \qps\ = $10^{7}$,
thus giving a $\sim$\,4\%\ probability of catching the
planet in its current state.  This probability is  
in line with the much smaller number of Jupiters below 1 d
compared to at 3--4 d.   

While the range \qps\ = 10$^{7}$--10$^{8}$
is much higher than values taken from binary stars, Penev \&\
Sasselov (2011) argue that, since stars are not spun up
by planetary-mass companions, the tidal forcing by planets 
does not resonate with stellar tides, and thus that dissipation is
much lower than in binary stars, and hence \qps\ can be
as high as 10$^{9}$.  

According to Matsumura \etal\ (2010) the obliquity would
be damped on a similar timescale to the orbital decay,
and the eccentricity would be damped on either the same or a 
faster timescale (depending on \qpp). Thus this scenario of 
significant orbital decay is consistent
with our finding of $\lambda$, $\psi$ and $e$ compatible with zero,
even if these were previously larger during the evolution
to and at 2\,$a_{\rm R}$. Winn \etal\ (2010) have argued that planets 
around cool stars with significant convection zones 
(\teff\ $<$ 6250 K) will rapidly evolve to aligned orbits, and
this fits with WASP-19's \teff\ of 5500 K.

Note that it is harder to find a plausible scenario if we start the
infall from larger orbital separations.  Given the steep dependence 
of timescale on
$a$ ($\tau \propto (a/R_{\rm *})^{5}$; Levrard \etal\ 2009), a small
enough \qps\ to fit within the age would then produce a very small
probability of finding it in its current orbit.  For this reason,
it is hard to explain WASP-19b's orbit by disc migration and 
tidal decay alone: if these processes were efficient enough to
have moved WASP-19b that far inwards, the same processes
would likely have destroyed it. 

The other possibility outlined by Guillochon \etal\ (2010) is 
that WASP-19b first migrated inwards to well within the ice line 
and was then scattered to its present semi-major axis.
This requires a lot of fine tuning to produce
$a$ as small as 1.2\,$a_{\rm R}$. The maximum radius at scattering 
can be only 0.01 of the ice line (Guillochon \etal\ 2010), the 
eccentricity after scattering must not be too high, and one 
has to avoid destruction by tidal orbital decay; yet if there
hasn't been significant tidal decay then one could expect
relics of the scattering in the form of eccentricity or misalignment.
Thus our findings of an aligned, circular orbit make this already 
unlikely scenario even less likely.

Thus, the current orbit of WASP-19b, the tightest of the known
exoplanets, is most plausibly explained if the planet first moved 
to near $\sim$\,2$a_{\rm R}$ by scattering
or by the Kozai mechanism, as has been
suggested for many of the hot Jupiters in the 3--4-d pileup, 
and circularised there; and
then, perhaps through starting at the lower edge of the pileup,
spiralled inward to its current location by orbital decay.

\acknowledgments
Based on observations made with the ESO 3.6-m/HARPS,  
program 084-C-0185, and the ESO NTT, and the Euler 1.2-m at
La Silla Observatory.


\begin{thebibliography}{}
\bibitem[Anderson et al.\ (2010)]{w19k}Anderson, D. R. et al., 2010, A\&A, 513, L3
\bibitem[Borucki et al.\ (2010]{kepler}Borucki,  W. J.,  et al., 2011, ApJ, submitted (arXiv:1102.0541) 
\bibitem[Bruntt  et al.\ (2010)]{bruntt}Bruntt, H. et al., 2010, MNRAS, 405, 1907
\bibitem[Claret (2000)]{claret00}Claret, A., 2000, A\&A, 363, 1081
\bibitem[Claret (2004)]{claret04}Claret, A., 2004, A\&A, 428, 1001
\bibitem[Enoch et al.\ (2010a)]{enoch}Enoch, B., Collier-Cameron, A., 
     Parley, N. R., \&\ Hebb, L., 2010a, A\&A, 516, 33. 
\bibitem[Fabrycky \& Tremaine (2007)]{fabrycky}Fabrycky, D. \&\ Tremaine, S., 2007, ApJ, 669, 1298
\bibitem[Fabrycky \& Winn (2009)]{fabwinn}Fabrycky, D. C. \&\ Winn, J. N., 2009, ApJ, 696, 1230
\bibitem[Ford \& Rasio (2006)]{fordrasio}Ford, E. B. \&\ Rasio, F. A., 2006, 
ApJ, 638, L45
\bibitem[Gibson et al.\ (2010)]{gibson}Gibson, N. P. et al., 2010, MNRAS, 404, L114
\bibitem[Guillochon et al.\ (2010)]{guill}Guillochon, J., Ramirez-Ruiz, E., Lin, D. N. C., 2010,  ApJ, submitted (arXiv:1012.2382)
\bibitem[Hebb et al.\ (2010)]{hebb}Hebb, L. et al., 2010, ApJ, 708, 224
\bibitem[Hellier et al.\ (2009)]{w18}Hellier, C. et al., 2009, Nature, 460, 1098
\bibitem[Hellier et al.\ (2010)]{hohp}Hellier, C. et al., 2010, Proceedings of the OHP conference ``Detection and dynamics of transiting exoplanets", (arXiv:1012.2286)
\bibitem[Hellier et al.\ (2011)]{w43}Hellier, C. et al., 2011, in preparation
\bibitem[Lin et al.\ (1996)]{lin}Lin, D. N. C., Bodenheimer, P., Richardson, D. C., 1996, Nature, 380, 606
\bibitem[Levrard et al.\ (2009)]{levrard}Levrard, B. Winisdoerffer, C. \&\ Chabrier, G., 2009, ApJ, 692, L9
\bibitem[Masumura et al.\ (2010)]{matsumura}Matsumura, S., Peale, S. J., Rasio, F. A., 2010, ApJ, 725, 1995
\bibitem[Morton et al.\ (2011)]{morton/johnson}Morton, T. D., Johnson, J. A., 
2011, ApJ submitted (arXiv:1010.4025) 
\bibitem[Nagasawa et al.\ (2008)]{nagasawa}Nagasawa, M., Ida, S., Bessho, T., 2008, ApJ, 678, 498
\bibitem[Naoz et al.\ (2010)]{naoz}Naoz, S., Farr, W. M., Lithwick, Y., Rasio, F. A., Teyssandier, J., 2010, arXiv:1011.2501
\bibitem[Narita et al.\ (2009)]{narita09}Narita, N., Sato, B., Hirano, T. \&\ Tamura, M., 2009, PASJ, 61, L35
\bibitem[Narita et al.\ (2010)]{narita10}Narita, N. et al., 2010, PASJ, submitted (arXiv:1008.3803) 
\bibitem[Penev \&\ Sasselov (2011)]{penev}Penev, K. \& Sasselov, D., 2011, ApJ, in press (arXiv:1102.3187) 
\bibitem[Rasio et al.\ (1996)]{rasio}Rasio, F. A., Tout, C. A., Lubow, S. H., 
     Livio, M., 1996,  ApJ, 470, 1187
\bibitem[Saar \&\ Donahue (1997)]{saar}Saar, S. H., \& Donahue, R. A., 1997, ApJ, 485, 319
\bibitem[Southworth et al.\ (2009)]{jkt}Southworth, J. et al., 2009, MNRAS, 396, 1023
\bibitem[Szab\'o \&\ Kiss (2011)]{szabo}Szab\'o, Gy. M., Kiss, L. L., 2011, ApJ, 727, L44
\bibitem[Triaud et al.\ (2009)]{triaud9}Triaud, A. H. M. J.,  2009, A\&A, 506, 377
\bibitem[Triaud et al.\ (2010)]{triaud10}Triaud, A. H. M. J.,  2010, A\&A, 524, 25
\bibitem[Triaud et al.\ (2011)]{triaud11}Triaud, A. H. M. J.,  2011, A\&A, submitted
\bibitem[Winn et al.\ (2009)]{winn09}Winn, J. N. et al.\ 2009, ApJ, 703, L99
\bibitem[Winn et al.\ (2010)]{winn10}Winn, J. N., Fabrycky, D., Albrecht, S., 
Johnson, J. A.,  2010, ApJ, 718, L145

\end{thebibliography}
\end{document}